\documentstyle[12pt]{article}
\begin{document}
\pagestyle{plain}
%  user-defined commands:
\newcommand{\be}{\begin{equation}}
\newcommand{\ee}{\end{equation}}
\newcommand{\bea}{\begin{eqnarray}}
\newcommand{\eea}{\end{eqnarray}}
\newcommand{\vp}{\varphi}
\newcommand{\pr}{\prime}
\newcommand{\sech} {{\rm sech}}
\newcommand{\cosech} {{\rm cosech}}
\newcommand{\psib} {\bar{\psi}}
\newcommand{\cosec} {{\rm cosec}}
\def\vs {\vskip .3 true cm}
\centerline {\bf N Fermion Ground State of Calogero-Sutherland type}
\centerline {\bf Models in Two and Higher Dimensions}
\vs
\centerline {\bf Avinash Khare$^*$}
\centerline {Institute of Physics, Sachivalaya Marg,}
\centerline {Bhubaneswar 751005, India.}
\vs
{\bf Abstract}

I obtain the exact ground state of $N$-fermions in $D$-dimensions 
$(D \geq 2)$ in 
case the $N$ particles are interacting via long-ranged two-body and three-body
interactions and further they are also interacting via the harmonic
oscillator potential. I also obtain the $N$-fermion ground state in case the 
oscillator potential is replaced by an $N$-body Coulomb-like interaction.
\vfill
e-mail : khare@beta.iopb.stpbh.soft.net
\eject

Recently there is a revival of interest in the area of exactly solvable
many body problems. A celebrated example of a solvable many-body system is
the well known Calogero-Sutherland (CS) model in one dimension [1-3] which
has found wide applications [4] in areas 
as diverse as quantum chaos and fractional 
statistics.In these models, one not only knows the exact $N$-boson as well as
$N$-fermion ground states but also the complete excitation spectrum. What
about two and higher 
dimensional many-body problems ? So far as I am aware off, the only
$N$-body problem 
which is exactly solvable in $D$-dimensions $(D \geq 2)$ is that
of $N$-bosons or $N$-fermions experiencing pairwise (or one body) harmonic
interaction. Apart from this example, there exist several $N$-body problems
[5-10] for which the $N$-boson ground state as well as radial excitation
spectrum over it is analytically known. In some of these cases, a class
of N-fermion excited states are also known analytically. However, to the best of
my knowledge, apart from the oscillator potential, there exist no other 
many-body problem in $D$-dimensions for which 
$N$-fermion ground state is exactly
known. 

In this note, I obtain the exact ground state of $N$-fermions 
in $D$-dimensions $(D \geq 2)$ in case the 
$N$-particles 
are interacting via long ranged two-body and three-body interactions
and 
further they are also interacting via an external harmonic oscillator potential.
As a by-product, I also discuss its relevance in the context of the
$N$-anyon ground state.
Finally, I also obtain the $N$-fermion
ground state in case the harmonic 
oscillator potential is replaced by an $N$-body
Coulomb-like potential [8]. 

Soon after the seminal work of Calogero [1], it was 
shown by Calogero and Marchioro [5]
that the $N$-boson 
ground state and radial excitations over it can be obtained in
the case of a three-dimensional $N$-body problem with two-body inverse square
interaction provided one also adds a 
long ranged three-body interaction which is 
not present 
in one dimension. Recently we [7] have generalized the Calogero-Marchioro
result to arbitrary number of dimensions and also considered the Sutherland
variant of the model. The model considered by us in $D$-dimensions 
is $(D \geq 2)$ [7]
\be\label{1}
H = - {1\over 2} \sum^N_{i=1} {\bf \nabla}^2_i 
+ g \sum^N_{i<j} {1\over r^2_{ij}} 
+ G \sum^N_{i<j,i\not = k, j \not = k} 
{{\bf r}_{ki} \cdot {\bf r}_{kj}\over r^2_{ki} r^2_{kj}}  
+ {1\over 2} \sum^N_{i=1} r^2_i
\ee
where we have set $\hbar = m = \omega = 1$ [11], $ {\bf r}_i$ 
is the position of the
i'th particle, ${\bf r}_{ij}= {\bf r}_i- {\bf r}_j$ 
denotes the relative separation
of the i'th and j'th particles, while $g$ and $G$ 
are the strengths of the two and
three-body interactions, respectively.

Unlike the $N$-boson ground state, it is in general far more difficult to obtain
the $N$-fermion ground state. 
This is because, in view of the Pauli exclusion principle,
the $N$-fermion ground state energy in $D$-dimensions, 
depends very sensitively
on the value of $N$ and $D$. 
As a result, even in the simplest case of $N$-fermions
in an oscillator potential, no 
general expression can be written for the ground state energy even 
though, for any given $N$ and $D$, one can immediately give its value. 

To obtain the $N$-fermion ground state of the system as given by eq. (1), 
we start with the ansatz
\be\label{2}
\psi = \prod\limits^{N}_{i<j} \mid{\bf r}_i - {\bf r}_j\mid^{\Lambda^f_D} 
\psi_1 \, .
\ee
On substituting this ansatz in the Schr\"odinger equation 
$H\psi=E\psi$ with $H$ as given
by eq. (1), it is easily shown that $\psi_1$ satisfies the equation
\be\label{3}
 - {1\over 2} \sum^N_{i=1} {\bf \nabla}^2_i \psi_1 
-\Lambda^f_D \sum^N_{i<j}  {({\bf r}_i - {\bf r}_j) \cdot 
({\bf \nabla}_i\psi_1 - {\bf \nabla}_j
\psi_1)\over r_{ij}^2}  + [ V - E + 2 \Lambda^f_D \sum^N_{i<j} {1\over
r_{ij}^2} ] \psi_1 = 0
\ee
provided $g$ and $G$ are related to $\Lambda^f_D$ by 
\be\label{4}
\Lambda^f_D = \sqrt G = {1\over 2} [\sqrt{D^2+4g} - D] \, .
\ee
Note that in this particular case $V = {1\over 2} \sum_i r^2_i$. Now, since we want the N-fermion
ground state, let us further substitute the ansatz
\be\label{5}
\psi_1 = \psi_S \phi (t\equiv \sum^N_{i=1} r^2_i)
\ee
where $\psi_S$ is the N x N Slater determinant which is the
$N$ free-fermion ground state state wave function. For example, one of the
free 3-fermion ground 
state wave function in $d\geq 3$ (while it is the unique one in D = 2) is 
given by
\be\label{6}
\psi^{N=3}_S = [ (x_1 y_2-x_2y_1)+(x_2y_3-x_3y_2)+(x_3y_1-x_1y_3)] \, .
\ee
On substituting the ansatz (5) in eq. (3) 
we find that $\phi$ satisfies the equation
\be\label{7}
t\phi^{''} (t) + \bigg [ {N(N-1)\over 2} \Lambda^f_D 
+e^f_0\bigg ] \phi'(t) +{1\over 2} (E -
V) \phi (t) = 0
\ee
where $e^f_0$ is 
the ground state energy (including the center of mass) of $N$-fermions
in an oscillator potential. 
For example, in the case of 3 particles ($N$=3), using eq. (6) it is easy
to verify eq. (7) and show that 
\be\label{8}
e^f_0 (N=3,D \geq 2 ) = {3D\over 2}+2 \, .
\ee
In the case of the oscillator potential $(V = {1\over 2} t)$, it is easily shown
that the exact solution to eq. (7) is
\be\label{9}
\phi (t) = e^{-t/2} \ L_n^{\Gamma^f_D}(t)
\ee
while the corresponding energy is
\be\label{10}
E^f_n = \bigg [ 2n+{N(N-1)\over 2}\Lambda^f_D +e^f_0 \bigg ]
\ee
which gives us the exact $N$-fermion ground state ($n$=0) as well as the radial 
excitation spectrum over it. Here $\Gamma^f_D$ is given by 
\be\label{11}
\Gamma^f_D = \bigg [{N(N-1)\over 2} \Lambda^f_D +e^f_0-1 \bigg ] \, .
\ee
Several comments are in order at this stage.
\begin{enumerate}

\item   It is worth noting that whereas the exact $N$-fermion ground state
is obtained when $g = G+D {\sqrt G}$ (see eq. (4)), the exact $N$-boson ground
state is obtained only in case $g = G +(D-2) {\sqrt G}$ [7]. 
The corresponding ground state energies and eigenfunctions are
\be\label{12}
E^f_0 = {N(N-1)\over 2} \Lambda^f_D + e^f_0
\ee
\be\label{12a}
\psi^f_0 = \prod\limits^{N}_{i< j} 
\mid{\bf r}_i - {\bf r}_j\mid^{\Lambda^f_D} \psi_S 
exp (-{1\over 2}\sum_i r^2_i)
\ee
\be\label{13}
E^b_0 = {N(N-1)\over 2} \Lambda^b_D + {ND\over 2}
\ee
\be\label{13a}
\psi^b_0 = \prod\limits^{N}_{i< j} 
\mid{\bf r}_i - {\bf r}_j\mid^{\Lambda^b_D}  
exp (-{1\over 2}\sum_{i=1}^N r^2_i) 
\ee
where
\be\label{13b}
\Lambda^b_D = {\sqrt G} = {1\over 2} [\sqrt {(D-2)^2 +4g} - (D-2)] \, .
\ee
Thus, as yet we do not
have a model (apart from the pure harmonic oscillator) 
in which the exact $N$-boson
as well as $N$-fermion ground state energies can be simultaneously obtained (for
the same value of the parameters) in a given dimension $D$.
Instead, what we find is that if for a given set of parameters, the ground
state energy of $N$-bosons in potential (1) 
can be obtained in $D$ dimensions, then for the
{\it same} set of parameters one can obtain the ground state energy of 
$N$-fermions but in $D-2$ dimensions. Conversely, if one can obtain the ground 
state energy of $N$-fermions in $D$-dimensions experiencing potential (1), then
for the same set of parameters, one can also obtain the ground state energy of
$N$-bosons in $D+2$ dimensions. It is worth pointing out that long time ago, a 
similar 
connection was obtained by Parisi and Sourlas [12] between the critical 
exponents of the random field model (which is one of the simplest disordered
system) in $D$-dimensions and the critical exponents of the same model in the 
absence of any disorder but in $D-2$ dimensions.
Similarly, they also showed [13] that the branched polymers in $D$-dimensions
belong to the same universality class as the Lee-Yang edge singularity in 
$D-2$ dimensions.
\item  The fact that eq. (12)  
indeed represents the ground state energy
of $N$-fermions can be shown as follows. 
Firstly, for $g = G = 0 = \Lambda^f_D$ (i.e. pure oscillator potential),
obviously $e^f_0$ is the ground state energy. Further, one can define the 
supersymmetric charges
\be\label{14}
Q_{x_i} = p_{x_i} - i x_i+i\Lambda^f_D \sum_{j(>i)} {x_i-x_j\over r_{ij}^2}
+i{\partial\over\partial x_i}(log \psi_S)
\ee
\be\label{15}
Q_{y_i} = p_{y_i} - i y_i+i\Lambda^f_D \sum_{j(>i)} {y_i-y_j\over r_{ij}^2}
+i{\partial\over\partial y_i}(log \psi_S)
\ee
etc. and their Hermitian conjugates and show that the H as given by eq. (1)
can be written as
\be\label{16}
H = {1\over 2} \sum^N_{i=1} (Q_{x_i}^+Q_{x_i}+Q^+_{y_i}Q_{u_i}+...)+E^F_0
\ee
where $E^f_0$ is as given by eq. (12) 
while $g$ and $G$ are related to
$\Lambda^f_D$ by eq. (4). 
Further, one can show that the $Q$'s annihilate the $N$-fermion
ground state as given by eq. (\ref{12a}).
Clearly, since the operator 
on the right hand side of eq. (\ref{16}) is positive definite
and it annihilates 
the ground state wave function, hence $E^f_0$ must be the $N$-fermion
ground state energy corresponding to the Hamiltonian (1). 

\item  For large $N$, how 
does the $N$-fermion ground state energy behaves as a function
of $N$ ? This is easily answered by 
noting that for large $N$, $e^f_0$ in $D$-dimensions
is given by
\be\label{17}
e^f_0 = {D\over D+1} (D!)^{1/D} N^{{D+1\over D}} + 0 (N^{{D-1\over D}}) \, .
\ee
Hence for any $D(\geq 2)$,   
and for large $N$, the ground state energy
of $N$-fermions goes like 
\be\label{18} 
E^f_0 \stackrel{N\rightarrow\infty}{\longrightarrow} {N^2\over 2} \Lambda^f_D 
+ {D\over D+1}
(D!)^{1/D} N^{{D+1\over D}} - {N\over 2}\Lambda^f_D+0(N^{{D-1\over D}})
\ee
In particular, note that for large $N$, 
the ground state energy of $N$-fermions
is maximum in $D$ = 2 and monotonically decreases with the number of dimensions.
On the other hand, the $N$-boson ground state
energy as given by eq. (\ref{13}) is least in two dimensions
and monotonically increases with the number of dimensions.

\item  The bosonic excited state spectrum is easily obtained by running 
through eqs. (2) to (9) putting $\psi_S = 1$ and replacing $\Lambda^f_D$
by $\Lambda^b_D$ as given by eq. (\ref{13b}). 
One finds that the bosonic radial eigenstates are given by 
\be\label {20}
E_n = \bigg [ 2n + {N(N-1)\over 2} \Lambda^b_D +{ND\over 2} \bigg ]
\ee
\be\label{21}
\psi_n = \prod\limits^N_{i<j} \mid {\bf r}_i- {\bf r}_j\mid^{\Lambda^b_D}
 exp ({-{1\over 2}\sum^N_{i=1}
r^2_i}) L^{\Gamma_D^b}_n
\ee
where
\be\label{22}
\Gamma^b_D = {N(N-1)\over 2}\Lambda^b_D + {ND\over 2}-1
\ee

I would now like to show that apart from the oscillator
potential, the exact $N$-fermion ground state energy can also be obtained
in case one replaces the oscillator potential in the Hamiltonian (1) by
the $N$-body potential 
\be\label{23}
V({\bf r}_1,{\bf r}_2,...,{\bf r}_N) = - {\alpha\over \sqrt{\sum_{i=1}^N r^2_i}}
\ee
To this end note that the steps as given by eqs. (2) to (7) are
independent of the form of the potential. On substituting
$t (= \sum_{i=1}^N r^2_i) = \rho^2$ in eq. (7), we find that in the case of the
potential (23) it takes the form
\be\label{24}
\phi^{''} (\rho) +{1\over\rho} [N(N-1)\Lambda^f_D+2 e^f_D-1] \phi'(\rho)+2
({\alpha\over \rho}-\mid E\mid )=0
\ee
where $e^f_0$ as before is the ground state energy of $N$-fermions in 
$D$-dimensions in an
oscillator potential.. It is easily shown shown that the solution to this
equation is 
\be\label{25}
\phi (\rho) = e^{-\sqrt{2\mid E\mid}\rho} 
L^{2\Gamma^f_D}_n (2\sqrt{2\mid E\mid}\rho)
\ee
\be\label{26}
E_n = - {\alpha^2\over 2\bigg [n+{N(N-1)\over 2}\Lambda^f_D
+ e^f_0 - {1\over 2} \bigg ]^2} \, , \ n = 0,1,2,...
\ee
where $\Gamma^f_D$ is as given by eq. (11). 
For $g = G = 0$ (i.e. $\Lambda^f_D = 0$),
this gives us the exact $N$-fermion ground state and radial excitations over it
in the potential (\ref{23}). This then provides another instance where exact
fermionic ground 
state can be obtained in the case of both the oscillator and the $N$-body
potential (\ref{23}) thereby providing one more support to the conjecture 
that whenever
an exact eigenstate can be 
obtained in the oscillator potential,  similar exact eigenstate can
also be obtained in the case of the $N$-body potential as given by 
(\ref{23}) [8].
\end{enumerate}

All these results are easily extended to the Calogero
variant i.e. by replacing the one body potential $\sum^N_{i=1} r^2_i$ by 
${1\over 2}\sum^N_{i<j}r^2_{ij}$ 
in Hamiltonian (1) as well as in eq. (\ref{23}). Everything
goes through except for the fact that now the energy is that of 
$N$-fermions minus
the center of mass energy (=$D$/2). 
For example, it is easily shown that the $N$-fermion
ground state and radial excitations over it are given by
\be\label{27}
\hat\psi_n = \prod\limits^N_{i<j} 
\mid{\bf r}_i-{\bf r}_j\mid^{\Lambda^f_D} \psi_S exp ( -{1\over 2}
{1\over\sqrt {2N}} \sum r^2_{ij})  
L_n^{\Gamma^f_D-{D\over 2}}({1\over\sqrt{2N}} \sum
r^2_{ij})
\ee
\be\label{28}
\hat E^f_n = 
 = \sqrt{{N\over 2}} \bigg [2n+{N(N-1)\over 2}\Lambda^f_D + e^f_0
-{D\over 2} \bigg ] \, .
\ee

It is worth pointing out that in the special case of three dimensions 
($D = 3$), Calogero and 
Marchioro [5]
had obtained 
these results long time ago [5] in case $N$ = 3,4. However, on comparing
their Eqs. (12) to (18) with our results, 
we find that there are misprints/mistakes in their
expressions. For example, 
according to them the 3-fermion energy is $\sqrt{{3\over 2}}
(2n+3\Lambda^f_{D=3}+4)$ while the 
correct expression should be $\sqrt{{3\over 2}}
(2n+3\Lambda^f_{D=3}+5)$ as seen from 
eq. (\ref{28}). The fact that our expression
is correct is easily seen by noting that in the limit of $g = G = 0$ (hence 
$\Lambda^f_D$ = 0), the ground state energy of 3-fermions (in an  
oscillator potential) in 3-dimensions is known to be 5 
(and not 4) after center
of mass has been taken out. 
Similarly, according to them $\Gamma^f_{D=3} (N=3) - {3\over 2}
= 3\Lambda^f_{D=3}+3$ while the correct expression is 
$3\Lambda^f_{D=3}+4$ as is easily verified by using eqs. (11) and (\ref{27}).
Similarly, their 4-fermion energy and eigenfunctions are not correct. As is
clear from eq. (\ref{28}), the correct expression for energy is 
$\sqrt 2 (2n+6\Lambda^F_{D=3}+15/2)$ and
not $\sqrt 2 (2n+6\Lambda^F_{D=3}+6)$ as claimed by them. Note that
the ground state energy of 4-fermions in 3-dimensions in pure oscillator
potential is 15/2 (and not 6) after the center of mass
has been taken out. Similarly, it may be noted that 
$\Gamma^f_{D=3}(N=4) - {3\over 2} = 6\Lambda^f_{D=3}+
{13\over 2}$ (see eq. (\ref{27})) 
and not $6\Lambda^f_{D=3}+5$ as claimed by them.

The results obtained in this paper have direct relevance to the problem of
$N$-anyons in an oscillator potential [10,14]. 
For example, one can now understand as to
why $N$-anyon linear state 
starting from the $N$-fermion ground has not been analytically
obtained so far. To see this, let us recall that the $N$-anyon 
Hamiltonian in the 
oscillator potential is related to the Hamiltonian (1) at 
$g = G (=\alpha^2$ say) by
\be\label{29}
H_{anyon} = H + \alpha \sum^N_{j > i=1} {l_{ij}\over r^2_{ij}}
\ee
where $l_{ij} = {\bf r}_{ij}\times {\bf p}_{ij}$, is the angular momentum
operator. However, as is clear from eqs. (\ref{13b}) and (4), at $g = G$, 
only
$N$-boson 
ground state can be obtained exactly but {\bf not} the $N$-fermion ground
state (it may be noted that anyons can only exist in two dimensions
). In 
this context it may also be noted that $N$-bosons, in their ground state,
always have zero angular momentum and hence the noninteracting $N$-anyon 
ground state starting from the 
$N$-boson ground state is identical with the interacting $N$-boson ground 
state in the Calogero-Marchioro model (with $g = G$).

What about the noninteracting $N$-anyon 
ground starting from the $N$-fermion ground state ? Using eq. (4)
it is clear that it cannot be a linear state since such a linear state is 
possible if $g\not = G$ while in the
anyon model $g$ is always equal to $G$. 

However, we can use the exact $N$-fermion ground state obtained in this paper 
to get
some information about the nature of the $N$-anyon ground state in an 
oscillator potential.
In particular, note that 
the N-anyon Hamiltonian is related to the Hamiltonian (1)
at $G = \alpha^2, g = \alpha^2+2\alpha$ by
\be\label{30}
H_{anyon} = H - \alpha \sum^N_{j > i=1} {l_{ij}\over r^2_{ij}} -2\alpha
\sum^N_{i<j} {1\over r^2_{ij}} \, .
\ee
Now what one could do is to treat second and third terms on the right hand side
of eq. (\ref{30}) as perturbation. In that case, to the zeroth order, there is
an exact linear 3-anyon state starting from the 3-fermion ground state. One now
has to calculate the effect of the last two terms by using perturbation theory
and study the nature of crossing with the exact $N$-anyon state starting from 
the $N$-boson ground state.
We hope to study this question in the near future.

{\bf Acknowledgement}

It is a pleasure to thank Somen Bhattacharjee for reminding me of the related 
work of 
Parisi and Sourlas.  

\vfill
\eject

{\bf Reference}

1. F. Calogero, J. Math. Phys. {\bf 10} (1969) 2191, 2197;  {\bf 12}
   (1971) 419.

2. B. Sutherland, J. Math. Phys. {\bf 12} (1971) 246.

3. M. Olshanetsky and A. Perelomov, Phys. Rev. {\bf 71} (1981)
   313;  {\bf 94} (1983) 313.

4. See for example the table given in 
   B. Simons, P. Lee and B. Altshuler, Phys. Rev. Lett. {\bf 72} (1994) 64
   and references their in.

5. F. Calogero and C. Marchioro, J. Math. Phys. {\bf 14} (1973) 182.

6. M.V.N. Murthy, R.K. Bhaduri and D. Sen, Phys. Rev. Lett {\bf 76}
(1996) 4103; R.K. Bhaduri, A. Khare, J. Law, M.V.N. Murthy and D. Sen,
J. Phys. {\bf A30} (1997) 257.

7. A. Khare and K. Ray, Phys. Lett. {\bf A230} (1997) 139.

8. A. Khare, Cond-Mat/9712113.

9. P.K. Ghosh, Phys. Lett. {\bf A229} (1996) 364.

10. For exact solutions for N anyons in an oscillator potential, 
see for example, A. Khare, {\it Fractional
Statistics and Quantum Theory }, Chapter 3, (World Scientific, Singapore, 1997) and references
their in.

11. Note that whereas in [7] we had set 2m=1, here we are setting m = 1
to be in conformity with the majority of the literature.

12. G. Parisi and N. Sourlas, Phys. Rev. Lett. {\bf 43} (1979) 744.

13. G. Parisi and N. Sourlas, Phys. Rev. Lett. {\bf 46} (1981) 871.

14. G. Date, M. Krishna and M.V.N. Murthy, Int. J. Mod. Phys. 
{\bf A9} (1994) 2545.
\end{document}